\documentclass[twocolumn,showpacs,floats,floatfix,superscriptaddress,aps,pra]{revtex4}
\usepackage{amsfonts}
\usepackage{amssymb}
\usepackage{amsmath}
\usepackage{graphicx}
\usepackage{bm}

\input{tcilatex}

\begin{document}

\author{P. A. Ivanov}
\affiliation{Department of Physics, Sofia University, James Bourchier 5 blvd, 1164 Sofia,
Bulgaria}
\author{E. S. Kyoseva}
\affiliation{Department of Physics, Sofia University, James Bourchier 5 blvd, 1164 Sofia,
Bulgaria}
\author{N. V. Vitanov}
\affiliation{Department of Physics, Sofia University, James Bourchier 5 blvd, 1164 Sofia,
Bulgaria}
\affiliation{Institute of Solid State Physics, Bulgarian Academy of Sciences,
Tsarigradsko chauss\'{e}e 72, 1784 Sofia, Bulgaria}
\title{Engineering of arbitrary U($N$) transformations by quantum
Householder reflections}
\date{\today }

\begin{abstract}
We propose a simple physical implementation of the quantum Householder
reflection (QHR) $\mathbf{M}(v)=\mathbf{I}-2\left\vert v\right\rangle
\left\langle v\right\vert $ in a quantum system of $N$ degenerate states
(forming a qunit) coupled simultaneously to an ancillary (excited) state by $%
N$ resonant or nearly resonant pulsed external fields. We also introduce the
generalized\emph{\ }QHR $\mathbf{M}(v;\varphi )=\mathbf{I}+\left(
e^{i\varphi }-1\right) \left\vert v\right\rangle \left\langle v\right\vert $%
, which can be produced in the same $N$-pod system when the fields are
appropriately detuned from resonance with the excited state. We use these
two operators as building blocks in constructing arbitrary preselected
unitary transformations. We show that the most general U($N$) transformation
can be factorized (and thereby produced) by either $N-1$ standard QHRs and
an $N$-dimensional phase gate, or $N-1$ generalized QHRs and a
one-dimensional phase gate. Viewed mathematically, these QHR factorizations
provide parametrizations of the U($N$) group. As an example, we propose a
recipe for constructing the quantum Fourier transform (QFT) by at most $N$
interaction steps. For example, QFT requires a single QHR for $N=2$, and
only two QHRs for $N=3$ and 4.
\end{abstract}

\maketitle


\section{Introduction}

Coherent control of quantum dynamics traditionally involves scenarios for
transfer of population, complete or partial, from one bound initial energy
state to another, single or superposition state, or a continuum of states.
Such techniques are well developed, particularly for two-state and
three-state systems, e.g. $\pi $ pulses \cite{Shore}, adiabatic passage
using one or more level crossings \cite{ARPC}, or stimulated Raman adiabatic
passage (STIRAP) and its extensions \cite{STIRAP}. Essentially all these
techniques start from a \emph{single} initial state; such a state can be
prepared experimentally, e.g. by optical pumping.

In the same time, in contemporary quantum physics implementations of
specific \emph{propagators} are often demanded, for example, some fields in
quantum information lean heavily on the quantum Fourier transform \cite{QI}.
Another example is quantum state engineering when a system starts in a
coherent \emph{superposition} of states; then one must construct the entire
propagator, while the above techniques provide only some transition
probabilities.

The implementation of such propagators is well understood and used for
qubits, i.e. two-state quantum systems, upon which the theory of quantum
information is primarily built \cite{QI}. On the other hand, qunits -- $N$%
-state quantum systems -- offer some advantages. For example, a \emph{qubit}
can encode two continuous parameters: the population ratio of the two qubit
states and the relative phase of their amplitudes.\ A \emph{qunit} in a pure
state can encode $2(N-1)$ parameters ($N-1$ populations and $N-1$ relative
phases), i.e. by using qunits information can be encoded in significantly
fewer particles than with qubits. This is beneficial for storing quantum
information, which can be particularly important if the number of particles
that can be used is restricted, e.g., due to decoherence \cite{QI}.
Furthermore, there are indications that using qunits can improve error
thresholds in fault tolerant computation.

Physical realizations of qunit operations in the existing proposals \cite%
{qudits-SU(2)}, however, are difficult to implement. These implementations
use sequences of U(2) operations, i.e. transformations acting at each
instance of time upon only two of the $N$ states of the qunit. The general U(%
$N$) transformation of a qunit requires $O(N^{2})$ such U(2) operations \cite%
{qudits-SU(2)}; hence the complexity increases rapidly with the qunit
dimension $N$, which makes qunit manipulations challenging, even for qutrits
($N=3$).

In this paper, we show that a general U($N$) transformation can be
implemented physically in a quantum system with \emph{only} $N$ interaction
steps. For this purpose we introduce a compact quantum implementation, in a 
\emph{single} interaction step, of the Householder reflection \cite%
{Householder}. The latter is a powerful and numerically very robust unitary
transformation, which has many applications in \emph{classical} data
analysis, e.g., in solving systems of linear algebraic equations, finding
eigenvalues of high-dimensional matrices, least-square optimization, QR
decomposition, etc. \cite{Householder applications}. The Householder
transformation, acting upon an arbitrary $N$-dimensional matrix, produces an
upper (or lower) triangular matrix by $N-1$ operations. When the initial
matrix is unitary, the resulting final matrix is diagonal, i.e. a phase gate
or a unit matrix. We use this propery to decompose an arbitrary U($N$)
matrix into Householder matrices and hence, design a recipe for physical
realization of a general U($N$) transformation.

The quantum Householder reflection (QHR) consists of a single interaction
step involving $N$ simultaneous pulsed fields. In contrast to the existing
U(2) realizations of qunit transformations, here each Householder reflection
acts simultaneously upon many states: $N$ states in the first step, $N-1$
states in the second, etc. This allows us to greatly reduce the number of
physical steps, from $O(N^{2})$ in U(2) realizations to only $O(N)$ in our
proposal.

We introduce two types of QHRs: standard QHR and generalized QHR; the latter
involves an additional phase factor. The physical realizations of both use
simultaneous pulses of precise areas in a system with an $N$-pod linkage
pattern, the difference being that the standard QHR operates on exact
resonance, whereas the generalized QHR requires specific detunings. Any
unitary matrix can be decomposed into $N-1$ standard QHRs and a phase gate,
or into $N$ generalized QHRs, without a phase gate. This advantage of the
generalized-QHR implementation derives from the additional phase in each
step, which delivers $N$ additional phases in the end, thereby making the
phase gate unnecessary.

This paper is organized as follows. In Sec. \ref{Sec-QHR} we define the
standard and generalized QHR gates and propose physical implementations. In
Sec. \ref{Sec-QHRdecomposition} we describe the decompositions of a general
U($N$) matrix by means of standard and generalized QHRs, which provide the
routes for realization of an arbitrary U($N$) transformation. In Sec. \ref%
{Sec-QFT} we apply these decompositions to quantum Fourier transforms. The
conclusions are summarized in Sec. \ref{Sec-conclusions}.

\section{Quantum Householder Reflection (QHR)\label{Sec-QHR}}

\subsection{Standart QHR}

An $N$-dimensional quantum Householder reflection (QHR) is defined as the
operator%
\begin{equation}
\mathbf{M}(v)=\mathbf{I}-2\left\vert v\right\rangle \left\langle
v\right\vert ,  \label{QHR}
\end{equation}%
where $\left\vert v\right\rangle $ is an $N$-dimensional normalized complex
column-vector and $\mathbf{I}$ is the identity operator. The QHR (\ref{QHR})
is hermitean and unitary, $\mathbf{M}(v)=\mathbf{M}(v)^{^{\dagger }}=\mathbf{%
M}(v)^{-1}$, which means that $\mathbf{M}(v)$ is involutary, $\mathbf{M}%
^{2}(v)=\mathbf{I}$; in addition, $\det \mathbf{M}(v)=-1$. If the vector $%
\left\vert v\right\rangle $ is real, $\mathbf{M}(v)$ has a simple geometric
interpretation: reflection with respect to an $(N-1)$-dimensional plane with
a normal vector $\left\vert v\right\rangle $; in the complex case the
interpretation is more involved. In general, the Householder vector $%
\left\vert v\right\rangle $ is complex, which implies that it contains $%
2(N-1)$ real parameters (taking into account the normalization condition and
the unimportant global phase).

\subsection{Generalized QHR}

We define the generalized QHR as%
\begin{equation}
\mathbf{M}(v;\varphi )=\mathbf{I}+\left( e^{i\varphi }-1\right) \left\vert
v\right\rangle \left\langle v\right\vert ,  \label{Householder}
\end{equation}%
where $\left\vert v\right\rangle $ is again an $N$-dimensional normalized
complex column-vector and $\varphi $ is an arbitrary phase. The standard QHR
(\ref{QHR}) is a special case of the generalized QHR (\ref{Householder}) for 
$\varphi =\pi $: $\mathbf{M}(v;\pi )\equiv \mathbf{M}(v)$. The generalized
QHR is unitary, 
\begin{equation}
\mathbf{M}(v;\varphi )^{-1}=\mathbf{M}(v;\varphi )^{\dagger }=\mathbf{M}%
(v;-\varphi ),  \label{M-unitary}
\end{equation}%
and its determinant is $\det \mathbf{M}=e^{i\varphi }$.

\subsection{Physical implementations}

\subsubsection{Coherently driven $N$-pod system}

%
%
%
%
%
%
%
%
%
\begin{figure}[tbp]
\includegraphics[width=60mm]{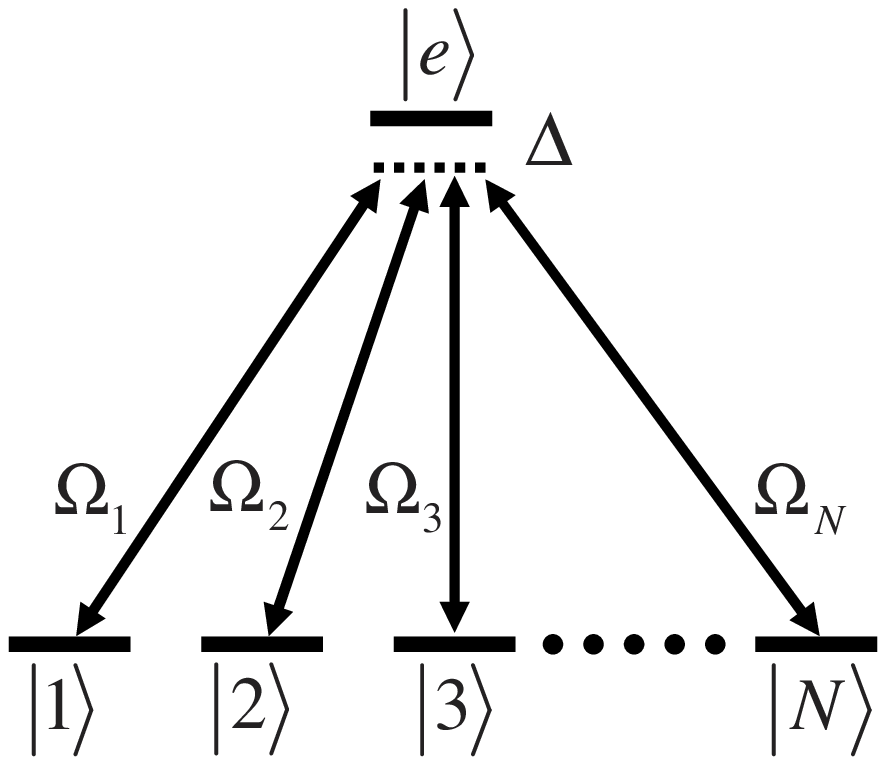}
\caption{Physical realization of the quantum Householder reflection: $N$
degenerate (in RWA sense) ground states, forming the \emph{qunit},
coherently coupled via a common excited state by pulsed external fields of
the same time dependence and the same detuning, but possibly different
amplitudes and phases. }
\label{Fig-Npod}
\end{figure}

The standard and generalized QHRs have simple physical realizations.
Consider the ($N+1$)-state system with $N$ degenerate [in the rotating-wave
approximation (RWA) sense \cite{Shore}] ground states $\left\vert
n\right\rangle $ ($n=1,2,\ldots ,N$), which represent the \emph{qunit},
coupled \emph{coherently} and \emph{simultaneously} by $N$ external fields
to an ancillary excited state $\left\vert e\right\rangle \equiv \left\vert
N+1\right\rangle $, as shown in Fig. \ref{Fig-Npod} \cite{Kyoseva}. Such an $%
N$-pod system can be formed, e.g., by coupling the magnetic sublevels of
several $J=1$ levels to a single $J=0$ level by polarized laser pulses \cite%
{Kyoseva}; for a qutrit only one $J=1$ level suffices. The propagator $%
U_{N+1}(t,t_{0})$ of this system obeys the Schr\"{o}dinger equation,%
\begin{equation}
i\hbar \frac{d}{dt}\mathbf{U}_{N+1}(t,t_{0})=\mathbf{H}(t)\mathbf{U}%
_{N+1}\left( t,t_{0}\right) ,  \label{SEq}
\end{equation}%
with the RWA Hamiltonian \cite{Shore}%
\begin{equation}
\mathbf{H}(t)=\frac{\hbar }{2}\left[ 
\begin{array}{ccccc}
0 & 0 & \cdots & 0 & \Omega _{1}\left( t\right) \\ 
0 & 0 & \cdots & 0 & \Omega _{2}\left( t\right) \\ 
\vdots & \vdots & \ddots & \vdots & \vdots \\ 
0 & 0 & \cdots & 0 & \Omega _{N}\left( t\right) \\ 
\Omega _{1}^{\ast }\left( t\right) & \Omega _{2}^{\ast }\left( t\right) & 
\cdots & \Omega _{N}^{\ast }\left( t\right) & 2\Delta \left( t\right)%
\end{array}%
\right] ,  \label{H}
\end{equation}%
and the initial condition $\mathbf{U}_{N+1}\left( t_{0},t_{0}\right) =%
\mathbf{I}$. The excited state $\left\vert e\right\rangle $ can be generally
off resonance by a detuning $\Delta \left( t\right) $ \cite{Kyoseva}, which,
however, must be the same for all fields. The functions $\Omega
_{1}(t),\ldots ,\Omega _{N}(t)$ are the Rabi frequencies of the couplings
between the ground states and the excited state; we require that they have
the same time dependence, described by the envelope function $f\left(
t\right) $, but we allow for different amplitudes $\chi _{n}$ and phases $%
\beta _{n}$, 
\begin{equation}
\Omega _{n}(t)=\chi _{n}f\left( t\right) e^{i\beta _{n}}\text{\quad }%
(n=1,2,\ldots ,N).  \label{Omega}
\end{equation}

By using the Morris-Shore transformation \cite{MS} the coupled ($N+1$)-state
system can be decomposed into a set of $N-1$ dark ground states, which are
superpositions of qunit states, and a two-state system, consisting of a
bright ground state and the excited state $\left\vert e\right\rangle $ \cite%
{Kyoseva}. This two-state system is driven by a Hamiltonian involving the
same detuning $\Delta (t)$ as in Eq. (\ref{H}), and the coupling is the
root-mean-square (rms) Rabi frequency $\Omega (t)=\sqrt{\sum_{n=1}^{N}\Omega
_{n}^{2}(t)}=\chi f(t)$.

The exact solution for the propagator reads \cite{Kyoseva}%
\begin{widetext}%
\begin{equation}
\mathbf{U}_{N+1}=\left[ 
\begin{array}{ccccc}
1+\left( a-1\right) \frac{\chi _{1}^{2}}{\chi ^{2}} & \left( a-1\right) 
\frac{\chi _{1}\chi _{2}e^{i\beta _{12}}}{\chi ^{2}} & \cdots  & \left(
a-1\right) \frac{\chi _{1}\chi _{N}e^{i\beta _{1N}}}{\chi ^{2}} & b\frac{%
\chi _{1}e^{i\beta _{1}}}{\chi } \\ 
\left( a-1\right) \frac{\chi _{1}\chi _{2}e^{-i\beta _{12}}}{\chi ^{2}} & 
1+\left( a-1\right) \frac{\chi _{2}^{2}}{\chi ^{2}} & \cdots  & \frac{\chi
_{2}\chi _{N}e^{i\beta _{2N}}}{\chi ^{2}} & b\frac{\chi _{2}e^{i\beta _{2}}}{%
\chi } \\ 
\vdots  & \vdots  & \ddots  & \vdots  & \vdots  \\ 
\left( a-1\right) \frac{\chi _{1}\chi _{N}e^{-i\beta _{1N}}}{\chi ^{2}} & 
\left( a-1\right) \frac{\chi _{2}\chi _{N}e^{-i\beta _{2N}}}{\chi ^{2}} & 
\cdots  & 1+\left( a-1\right) \frac{\chi _{N}^{2}}{\chi ^{2}} & b\frac{\chi
_{N}e^{i\beta _{N}}}{\chi } \\ 
-b^{\ast }\frac{\chi _{1}e^{-i\beta _{1}}}{\chi } & -b^{\ast }\frac{\chi
_{2}e^{-i\beta _{2}}}{\chi } & \cdots  & -b^{\ast }\frac{\chi _{N}e^{-i\beta
_{N}}}{\chi } & a^{\ast }%
\end{array}%
\right].   \label{Un1}
\end{equation}%
\end{widetext}Here $\chi =\sqrt{\sum_{n=1}^{N}\chi _{n}^{2}}$ is the rms
peak Rabi frequency and $\beta _{km}=\beta _{k}-\beta _{m}$ $\left(
k,m=1,2,\ldots ,N\right) $ are the relative phases of the external fields.
The complex parameters $a$ and $b$ (with $\left\vert b\right\vert
^{2}=1-\left\vert a\right\vert ^{2}$) are the Cayley-Klein parameters of the
SU(2) propagator for the Morris-Shore bright-excited two-state system.

\subsubsection{Standart QHR: exact resonance}

In the case of exact resonance ($\Delta =0$) the Cayley-Klein parameters for
any pulse shape $f(t)$ are 
\begin{equation}
a=\cos \frac{A}{2},\quad b=-i\sin \frac{A}{2},  \label{resonance}
\end{equation}%
where $A$ is the rms pulse area, 
\begin{equation}
A=\chi \int_{t_{i}}^{t_{f}}f\left( t\right) dt.  \label{rms area}
\end{equation}%
If 
\begin{equation}
A=2\left( 2k+1\right) \pi \text{ \ \ }\left( k=0,1,2,\ldots \right) ,
\label{area}
\end{equation}%
then $a=-1$, $\sin \left( A/2\right) =0$, and the last row and column of the
propagator (\ref{Un1}) vanish, except for the diagonal element, which is $-1$%
; the propagator (\ref{Un1}) reduces to%
\begin{equation}
\mathbf{U}_{N+1}=\left[ 
\begin{array}{cccc}
\ulcorner &  & \urcorner & 0 \\ 
& \mathbf{U}^{\pi } &  & \vdots \\ 
\llcorner &  & \lrcorner & 0 \\ 
0 & \cdots & 0 & -1%
\end{array}%
\right] .  \label{UN}
\end{equation}%
Here $\mathbf{U}^{\pi }$ is an $N$-dimensional unitary matrix (with $\det 
\mathbf{U}^{\pi }=-1$),\emph{\ }which represents the propagator within the $%
N $-state degenerate manifold; it has exactly the QHR form (\ref{QHR}), $%
\mathbf{U}^{\pi }=\mathbf{M}(v;\pi )=\mathbf{M}(v)$. The components of the $%
N $-dimensional QHR vector $\left\vert v\right\rangle $ are the normalized
Rabi frequencies, with the accompanying phases,%
\begin{equation}
\left\vert v\right\rangle =\frac{1}{\chi }\left[ \chi _{1}e^{i\beta
_{1}},\chi _{2}e^{i\beta _{2}},\ldots ,\chi _{N}e^{i\beta _{N}}\right] ^{T}.
\label{V}
\end{equation}%
Hence the propagator $\mathbf{U}^{\pi }$ within the degenerate $N$-state
manifold of the $N$-pod system driven by the Hamiltonian (\ref{H}), with $%
\Delta =0$ and rms pulse area (\ref{rms area}), represents indeed a \emph{%
physical realization of QHR in a single interaction step}. Any QHR vector (%
\ref{V}) can be produced by appropriately selecting the peak couplings $\chi
_{n}$ and the phases $\beta _{n}$, while obeyng Eq. (\ref{area}) (e.g., by
adjusting the pulse duration).

\subsubsection{Generalized QHR}

The unitary propagator (\ref{Un1}) for $a=e^{i\varphi }$ ($\left\vert
b\right\vert =0$) reduces to%
\begin{equation}
\mathbf{U}_{N+1}=\left[ 
\begin{array}{cccc}
\ulcorner &  & \urcorner & 0 \\ 
& \mathbf{U}^{\varphi } &  & \vdots \\ 
\llcorner &  & \lrcorner & 0 \\ 
0 & \cdots & 0 & e^{-i\varphi }%
\end{array}%
\right] ,  \label{UNRZ}
\end{equation}%
where, as is easily verified, we have $\mathbf{U}^{\varphi }=\mathbf{M}%
(v;\varphi )$, and hence, the propagator $\mathbf{U}^{\varphi }$ represents
a \emph{physical realization of the generalized QHR} (\ref{Householder}).
The vector $\left\vert v\right\rangle $ is again given by Eq. (\ref{V}). The
condition $a=e^{i\varphi }$ for $\varphi \not=0,\pi $ can only be realized
off resonance ($\Delta \not=0$). There is a beautiful off-resonance solution
to the Schr\"{o}dinger equation -- the Rosen-Zener (RZ) model -- which we
shall use here to exemplify the generalized QHR.

The Rozen-Zener (RZ) model \cite{RZ} can be seen as an extension of the
resonance solution (\ref{resonance}) to nonzero detuning for a special pulse
shape (hyperbolic-secant), 
\begin{subequations}
\label{RZ model}
\begin{eqnarray}
f\left( t\right) &=&\text{sech}\left( t/T\right) ,  \label{pshape} \\
\Delta \left( t\right) &=&\Delta _{0}.  \label{detuning}
\end{eqnarray}%
The Cayley-Klein parameter $a$ reads \cite{RZ,Kyoseva} 
\end{subequations}
\begin{equation}
a=\frac{\Gamma ^{2}\left( \frac{1}{2}+\frac{1}{2}i\Delta _{0}T\right) }{%
\Gamma \left( \frac{1}{2}+\frac{1}{2}\chi T+\frac{1}{2}i\Delta _{0}T\right)
\Gamma \left( \frac{1}{2}-\frac{1}{2}\chi T+\frac{1}{2}i\Delta _{0}T\right) }%
,  \label{a-rz}
\end{equation}%
where $\Gamma (z)$ is Euler's gamma function. Using the reflection formula $%
\Gamma \left( \frac{1}{2}+z\right) \Gamma \left( \frac{1}{2}-z\right) =\pi
/\cos \pi z$, we find 
\begin{equation}
\left\vert a\right\vert ^{2}=1-\frac{\sin ^{2}\left( \frac{1}{2}\pi \chi
T\right) }{\cosh ^{2}\left( \frac{1}{2}\pi \Delta _{0}T\right) }.
\label{modula}
\end{equation}

Hence in this model, $\left\vert a\right\vert =1$ for $\chi T=2l$ $\left(
l=0,1,2,\ldots \right) $; then the last row and the last column of the
propagator (\ref{Un1}) vanish, except the diagonal element. The phase $%
\varphi $ of $a=e^{i\varphi }$ depends on the detuning $\Delta _{0}$ and for
an arbitrary integer $l$ we find from Eq. (\ref{a-rz})%
\begin{equation}
a=e^{i\varphi }=\prod_{k=0}^{l-1}\frac{\Delta _{0}T+i\left( 2k+1\right) }{%
\Delta _{0}T-i\left( 2k+1\right) },  \label{phasea}
\end{equation}%
and hence%
\begin{equation}
\varphi =2\arg \prod_{k=0}^{l-1}\left[ \Delta _{0}T+i\left( 2k+1\right) %
\right] .  \label{phi}
\end{equation}%
This can be seen as an algebraic equation for $\Delta _{0}$, which has $l$
real solutions. For example, for $l=1$ [which corresponds to rms pulse area $%
A=2\pi $], we have $\Delta _{0}T=\cot \left( \varphi /2\right) $. Hence the
generalized-QHR phase $\varphi $ can be produced by an appropriate choice of
the detuning $\Delta _{0}$.

The use of nonresonant interaction, besides providing an additional phase
parameter, has another important advantage over resonant pulses: \emph{lower
transient population} of the intermediate state. This can be crucial if the
lifetime of this state is short compared to the interaction duration.
Equation (\ref{phi}) provides the opportunity to control this transient
population, which is proportional to $\Delta ^{-2}$, by using large peak
Rabi frequency (implying larger $l$) and find the largest solution for $%
\Delta $. It is important that the \emph{standard} QHR can also be realized
off resonance, by selecting a detuning $\Delta _{0}$ for which $\varphi =\pi 
$.

\section{QHR decomposition of U($N$)\label{Sec-QHRdecomposition}}

\subsection{Standard-QHR decomposition}

We shall show that QHR is a very efficient tool for constructing a general U(%
$N$) qunit gate. In particular, we shall show that any $N$-dimensional
unitary matrix $\mathbf{U}$ ($\mathbf{U}^{-1}=\mathbf{U}^{\dag }$) can be
expressed as a product of $N-1$ standard QHRs $\mathbf{M}(v_{n})$ ($%
n=1,2,...,N-1$) and a phase gate $\mathbf{\Phi }\left( \phi _{1},\phi
_{2},\ldots ,\phi _{N}\right) $,\emph{\ }%
\begin{equation}
\mathbf{U}=\mathbf{M}(v_{1})\mathbf{M}(v_{2})\cdots \mathbf{M}(v_{N-1})%
\mathbf{\Phi }\left( \phi _{1},\phi _{2},\ldots ,\phi _{N}\right) ,
\label{SU(N) decomposition standard}
\end{equation}%
where 
\begin{equation}
\mathbf{\Phi }\left( \phi _{1},\phi _{2},\ldots ,\phi _{N}\right) =\text{diag%
}(e^{i\phi _{1}},e^{i\phi _{2}},\ldots ,e^{i\phi _{N}}).  \label{F}
\end{equation}

We shall prove this assertion by explicitly constructing the decomposition (%
\ref{SU(N) decomposition standard}). The standard QHRs $\mathbf{M}(v_{n})$
involve vectors $\left\vert v_{n}\right\rangle $, which we construct as
follows. First we define the normalized vector $\left\vert
v_{1}\right\rangle $ as 
\begin{equation}
\left\vert v_{1}\right\rangle =\frac{\left\vert u_{1}\right\rangle -e^{i\phi
_{1}}\left\vert e_{1}\right\rangle }{\sqrt{2\left[ 1-\text{Re}\left(
u_{11}e^{-i\phi _{1}}\right) \right] }},
\end{equation}%
where the vector $\left\vert u_{n}\right\rangle $ denotes the $n$th column
of $\mathbf{U}=\left\{ u_{kn}\right\} $, $\phi _{1}=\arg u_{11}$, and $%
\left\vert e_{1}\right\rangle =\left[ 1,0,...,0\right] ^{T}$. We find 
\begin{subequations}
\begin{eqnarray}
\mathbf{M}(v_{1})\left\vert u_{1}\right\rangle &=&e^{i\phi _{1}}\left\vert
e_{1}\right\rangle ,  \label{M-column} \\
\mathbf{M}(v_{1})\left\vert u_{n}\right\rangle &=&\left\vert
u_{n}\right\rangle +2e^{-i\phi _{1}}u_{1n}\left\vert v_{1}\right\rangle ,
\label{M-row} \\
\left\langle e_{1}\right\vert \mathbf{M}(v_{1})\left\vert u_{n}\right\rangle
&=&0\text{\quad }\left( n=2,3,\ldots ,N\right) .  \label{M-row2}
\end{eqnarray}%
Hence the action of $\mathbf{M}(v_{1})$ upon $\mathbf{U}$ nullifies the
first row and the first column except for the first element, 
\end{subequations}
\begin{equation}
\mathbf{M}(v_{1})\mathbf{U}=\left[ 
\begin{array}{cccc}
e^{i\phi _{1}} & 0 & \cdots & 0 \\ 
0 & \ulcorner &  & \urcorner \\ 
\vdots &  & \mathbf{U}_{N-1} &  \\ 
0 & \llcorner &  & \lrcorner%
\end{array}%
\right] ,
\end{equation}%
where $\mathbf{U}_{N-1}$ is a U($N-1$) matrix. We repeat the same procedure
on $\mathbf{M}(v_{1})\mathbf{U}$ and construct the vector $\left\vert
v_{2}\right\rangle $, 
\begin{equation}
\left\vert v_{2}\right\rangle =\frac{\left\vert u_{2}^{\prime }\right\rangle
-e^{i\phi _{2}}\left\vert e_{2}\right\rangle }{\sqrt{2\left[ 1-\text{Re}%
\left( u_{22}^{\prime }e^{-i\phi _{2}}\right) \right] }},
\end{equation}%
where the vector $\left\vert u_{2}^{\prime }\right\rangle $ is the second
column of $\mathbf{M}(v_{1})\mathbf{U}$, $\phi _{2}=\arg \left[ \mathbf{M}%
(v_{1})\mathbf{U}\right] _{22}$, and $\left\vert e_{2}\right\rangle =\left[
0,1,0,\ldots ,0\right] ^{T}$.\emph{\ } The corresponding QHR $\mathbf{M}%
(v_{2})$, applied to $\mathbf{M}(v_{1})\mathbf{U}$, has the following
effects: (i) nullifies the second row and the second column of $\mathbf{M}%
\mathbf{(}v_{1})\mathbf{U}$ except for the diagonal element, which becomes $%
e^{i\phi _{2}}$, and (ii) does not change the first row and the first
column. By repeating the same procedure $N-1$ times, we construct $N-1$
consecutive Householder reflections, which nullify all off-diagonal
elements, to produce a diagonal matrix comprising $N$ phase factors, 
\begin{equation}
\mathbf{M(}v_{N-1})\cdots \mathbf{M(}v_{1})\mathbf{U}=\mathbf{\Phi }(\phi
_{1},\phi _{2},\ldots ,\phi _{N}),
\end{equation}%
which completes the proof of Eq. (\ref{SU(N) decomposition standard}) since $%
\mathbf{M(}v)=\mathbf{M(}v)^{-1}$. If $\mathbf{U}$ is a SU($N$) matrix then $%
\det \mathbf{\Phi }=\pm 1$, meaning $\sum_{n=1}^{N}\phi _{n}=0$ or $\pi $.

We note that the choice of the QHRs $\mathbf{M}(v_{n})$ is not unique; for
example, the first QHR $\mathbf{M}(v_{1})$\ can be constructed from the
first row of $\mathbf{U}$, instead of the first column. Furthermore, the
final diagonal matrix (\ref{F}) occurs due to the unitarity of $\mathbf{U}$,
which leads to Eq. (\ref{M-row2}); a QHR sequence produces a triangular
matrix in general.

The QHR decomposition (\ref{SU(N) decomposition standard}) of the U($N$)
group into $N-1$ Householder matrices (\ref{QHR}) and a phase gate provides
a simple and efficient physical realization of a general transformation of a
qunit by only $N-1$ interaction steps and a phase gate; this is a
significant advance compared to $O(N^{2})$ operations in existing recipes.
Each QHR vector is $N$-dimensional, but the nonzero elements decrease from $%
N $ in $\left\vert v_{1}\right\rangle $ to just 2 in $\left\vert
v_{N-1}\right\rangle $, and so does the number of fields required for each
QHR, see Eq. (\ref{V}).

The decomposition (\ref{SU(N) decomposition standard}) is also of
mathematical interest because it provides a very natural parametrization of
the U($N$) group. Indeed, a QHR vector with $n$ nonzero elements contains $%
2(n-1)$ real parameters (because of the normalization and the irrelevant
global phase). The phase gate (\ref{F}) contains $N$ phases. Hence Eq. (\ref%
{SU(N) decomposition standard}) involves $\sum_{n=2}^{N}2(n-1)+N=N^{2}$ real
parameters, as should be the case for a general U($N$) matrix.

\subsection{Generalized-QHR decomposition}

We shall show now that any unitary matrix $\mathbf{U}$ can be expressed as a
product of $N$ \emph{generalized} QHRs $\mathbf{M}\left( v_{n};\varphi
_{n}\right) $ $\left( n=1,2,\ldots ,N\right) $ defined by Eq. (\ref%
{Householder}), \emph{without a phase gate}, that is%
\begin{equation}
\mathbf{U}=\prod_{n=1}^{N}\mathbf{M}(v_{n};\varphi _{n}).
\label{SU(N) decomposition}
\end{equation}

We first define the normalized vector 
\begin{equation}
\left\vert v_{1}\right\rangle =\frac{1}{e^{-i\varphi _{1}}-1}\sqrt{\frac{%
2\sin \left( \varphi _{1}/2\right) }{\left\vert 1-u_{11}\right\vert }}\left(
\left\vert u_{1}\right\rangle -\left\vert e_{1}\right\rangle \right) ,
\end{equation}%
where the vector $\left\vert u_{n}\right\rangle $ denotes again the $n$th
column of $\mathbf{U}$ and $\varphi _{1}=2\arg \left( 1-u_{11}\right) -\pi $%
. It is readily seen that 
\begin{subequations}
\label{M}
\begin{eqnarray}
\mathbf{M}(v_{1};-\varphi _{1})\left\vert u_{1}\right\rangle &=&\left\vert
e_{1}\right\rangle ,  \label{M1-column} \\
\left\langle e_{1}\right\vert \mathbf{M}(v_{1};-\varphi _{1})\left\vert
u_{n}\right\rangle &=&0\text{\quad }(n=2,3,\ldots ,N\ ).  \label{M1-row}
\end{eqnarray}%
Therefore, the action of $\mathbf{M}(v_{1};-\varphi _{1})$ upon $\mathbf{U}$
nullifies the first row and the first column except for the first element,
which is turned into unity, 
\end{subequations}
\begin{equation}
\mathbf{M}(v_{1};-\varphi _{1})\mathbf{U}=\left[ 
\begin{array}{cccc}
1 & 0 & \cdots & 0 \\ 
0 & \ulcorner &  & \urcorner \\ 
\vdots &  & \mathbf{U}_{N-1} &  \\ 
0 & \llcorner &  & \lrcorner%
\end{array}%
\right] ,  \label{M1U}
\end{equation}%
where $\mathbf{U}_{N-1}$ is a U($N-1$) matrix. We repeat the same procedure
on $\mathbf{U}_{N-1}$ and construct the vector 
\begin{equation}
\left\vert v_{2}\right\rangle =\frac{1}{e^{-i\varphi _{2}}-1}\sqrt{\frac{%
2\sin \left( \varphi _{2}/2\right) }{\left\vert 1-u_{22}^{\prime
}\right\vert }}\left( \left\vert u_{2}^{\prime }\right\rangle -\left\vert
e_{2}\right\rangle \right) ,
\end{equation}%
where the vector $\left\vert u_{2}^{\prime }\right\rangle $ is the second
column of $\mathbf{M}(v_{1};-\varphi _{1})\mathbf{U}$ and $\varphi
_{2}=2\arg \left( 1-u_{22}^{\prime }\right) -\pi $. The action of $\mathbf{M}%
(v_{2};-\varphi _{2})$ upon $\mathbf{M}(v_{1};-\varphi _{1})\mathbf{U}$ has
the following effects: (i) nullifies the second row and the second column of 
$\mathbf{M}(v_{1};-\varphi _{1})\mathbf{U}$ except for the diagonal element
which is turned into unity, and (ii) does not change the first row and the
first column of $\mathbf{M}(v_{1};-\varphi _{1})\mathbf{U}$. By repeating
the same procedure $N$ times, we construct $N$ consecutive generalized
Householder reflections, which nullify all off-diagonal elements to produce
the identity matrix, 
\begin{equation}
\prod_{n=N}^{1}\mathbf{M}(v_{n};-\varphi _{n})\mathbf{U}=\mathbf{I}\text{.}
\label{identity}
\end{equation}%
By recalling Eq. (\ref{M-unitary}) we obtain Eq. (\ref{SU(N) decomposition})
immediately. Note that the last QHR $\mathbf{M}(v_{N};\varphi _{N})=\mathbf{%
\Phi }(0,\ldots ,0,\varphi _{N})$ is actually a one-dimensional phase gate.

Therefore the use of generalized QHRs replaces the $N$-dimensional phase
gate needed in the standard-QHR implementation (\ref{SU(N) decomposition
standard}) by a one-dimensional phase gate $\mathbf{\Phi }(0,\ldots
,0,\varphi _{N})$. We point out that again, as for the standard QHRs $%
\mathbf{M}(v_{n})$, the choice of any of the generalized QHRs $\mathbf{M}%
(v_{n};\varphi _{n})$ is not unique because it can be constructed from the
respective row, rather than the column, of the corresponding matrix.

\subsection{Examples}

\subsubsection{Qubit}

As an example of the QHR decomposition we first consider the qubit, which is
the conventional system for quantum information processing. The conventional
realization of a general U(2) transformation involves three interactions:
two phase gates and one rotation $\mathbf{R}\left( \vartheta \right) $ \cite%
{QI},%
\begin{equation}
\mathbf{U}=\mathbf{\Phi }\left( \alpha _{1},\alpha _{2}\right) \mathbf{R}%
\left( \vartheta \right) \mathbf{\Phi }\left( 0,\alpha _{3}\right) .
\label{SU(2)}
\end{equation}

Already for a qubit, the QHR implementations (\ref{SU(N) decomposition
standard}) and (\ref{SU(N) decomposition}) are superior to Eq. (\ref{SU(2)})
because they only require one QHR and one phase gate, 
\begin{subequations}
\label{SU(2) implementation}
\begin{eqnarray}
\mathbf{U} &=&\mathbf{M}(v)\mathbf{\Phi }(\phi _{1},\phi _{2}),
\label{SU(2) standard} \\
\mathbf{U} &=&\mathbf{M}(v;\varphi _{1})\mathbf{\Phi }\left( 0,\varphi
_{2}\right) .  \label{SU(2) generalized}
\end{eqnarray}

\subsubsection{Qutrit}

As a second example we consider a qutrit --- a three-state quantum system.
The most general transformation of a qutrit belongs to the U(3) group, which
can be parametrized by nine real parameters; respectively, the SU(3) group
is described by eight real parameters. A SU(2) factorization of SU(3) reads 
\cite{SU(3)} 
\end{subequations}
\begin{equation}
\mathbf{U}=\mathbf{R}_{23}\left( \alpha _{1},\beta _{1},\gamma _{1}\right) 
\mathbf{R}_{12}\left( \alpha _{2},\beta _{2},\alpha _{2}\right) \mathbf{R}%
_{23}\left( \alpha _{3},\beta _{3},\gamma _{3}\right) ,  \label{SU(3)}
\end{equation}%
where $R_{mn}$ are SU(2) subgroups of SU(3), with the SU(2) submatix
occupying the $m$th and $n$th rows and columns of $R_{mn}$. Hence this
implementation (\ref{SU(3)}) of SU(3) requires three SU(2) gates, each
involving three qubit gates (\ref{SU(2)}), i.e. \emph{nine} qubit gates in
total (which can be reduced to \emph{seven} by combining adjacent phase
gates). With the present QHR implementation (\ref{SU(2) implementation}) of
SU(2) the number of operations can be reduced to \emph{six}.

Already for SU(3) or U(3), the present QHR implementations (\ref{SU(N)
decomposition standard}) and (\ref{SU(N) decomposition}) are considerably
more efficient because they require \emph{only two} QHRs and a phase gate, 
\begin{subequations}
\label{SU(3) implementation}
\begin{eqnarray}
\mathbf{U} &=&\mathbf{M}(v_{1})\mathbf{M}(v_{2})\mathbf{\Phi }(\phi
_{1},\phi _{2},\phi _{3}),  \label{SU(3) standard} \\
\mathbf{U} &=&\mathbf{M}(v_{1};\varphi _{1})\mathbf{M}(v_{2};\varphi _{2})%
\mathbf{\Phi }(0,0,\varphi _{3}).  \label{SU(3) generalized}
\end{eqnarray}

As an example, the arbitrarily chosen SU(3) gate 
\end{subequations}
\begin{equation}
\mathbf{U}=\left[ 
\begin{array}{ccc}
0.864e^{-2\pi i/3} & 0.282e^{15\pi i/19} & 0.416e^{-7\pi i/8} \\ 
0.382e^{0.140\pi i} & 0.902e^{7\pi i/11} & 0.203e^{0.808\pi i} \\ 
0.327e^{-0.789\pi i} & 0.328e^{4\pi i/5} & 0.886e^{0.035\pi i}%
\end{array}%
\right]  \label{SU(3)-example}
\end{equation}%
(keeping 3 significant digits) can be realized with two standard QHRs and a
phase gate, with 
\begin{subequations}
\begin{eqnarray}
\left\vert v_{1}\right\rangle &=&[0.260e^{i\pi /3},0.734e^{0.140\pi
i},0.628e^{-0.789\pi i}]^{T}, \\
\left\vert v_{2}\right\rangle &=&[0,0.651e^{-0.134\pi i},0.759e^{0.710\pi
i}]^{T}, \\
\mathbf{\Phi } &=&\text{diag}\left\{ e^{-0.667\pi i},e^{0.866\pi
i},e^{-0.199\pi i}\right\} .
\end{eqnarray}%
Alternatively, the same SU(3) gate (\ref{SU(3)-example}) can be realized by
two generalized QHRs and a phase gate (\ref{SU(3) generalized}), with $%
\varphi _{1}=-0.693\pi $, $\varphi _{2}=0.653\pi $, $\varphi _{3}=0.04\pi $,
and 
\end{subequations}
\begin{subequations}
\begin{eqnarray}
\left\vert v_{1}\right\rangle &=&\left[ 0.955e^{0.307\pi
i},0.226e^{-0.707\pi i},0.193e^{0.364\pi i}\right] ^{T}, \\
\left\vert v_{2}\right\rangle &=&\left[ 0,0.987e^{0.347\pi
i},0.161e^{-0.383\pi i}\right] ^{T}.
\end{eqnarray}

\section{Quantum Fourier transform}

The quantum Fourier transform (QFT) is a key ingredient in quantum
factoring, quantum search, generalized phase estimation, the hidden subgroup
problem, and many other quantum algorithms \cite{QI}. The QFT is defined as
the unitary operator with the following action on an orthonormal set of
states $\left\vert n\right\rangle $ $(n=1,2\ldots ,N)$: 
\end{subequations}
\begin{equation}
\mathbf{U}_{N}^{F}\left\vert n\right\rangle =\frac{1}{\sqrt{N}}%
\sum_{k=1}^{N}e^{2\pi i(n-1)(k-1)/N}\left\vert k\right\rangle .  \label{QFT}
\end{equation}

\subsection{Qubit}

For a qubit, $\mathbf{U}^{F}$ is the Hadamard gate \cite{QI},%
\begin{equation}
\mathbf{U}_{2}^{F}=\frac{1}{\sqrt{2}}\left[ 
\begin{array}{cc}
1 & 1 \\ 
1 & -1%
\end{array}%
\right] ,  \label{QFT2}
\end{equation}%
which can be written as a single QHR, $\mathbf{U}_{2}^{F}=\mathbf{M}(v)$,
with 
\begin{equation}
\left\vert v\right\rangle =\frac{1}{2}\left[ -\sqrt{2-\sqrt{2}},\sqrt{2+%
\sqrt{2}}\right] ^{T}.  \label{V-2}
\end{equation}
Here the standard and generalized QHRs coincide.

\subsection{Qutrit}

For a qutrit the QFT matrix reads%
\begin{equation}
\mathbf{U}_{3}^{F}=\frac{1}{\sqrt{3}}\left[ 
\begin{array}{ccc}
1 & 1 & 1 \\ 
1 & e^{2\pi i/3} & e^{-2\pi i/3} \\ 
1 & e^{-2\pi i/3} & e^{2\pi i/3}%
\end{array}%
\right] .  \label{QFT3}
\end{equation}%
The \emph{standard}-QHR decomposition reads 
\begin{subequations}
\label{QFT3 standard}
\begin{eqnarray}
\mathbf{U}_{3}^{F} &=&\mathbf{M}(v_{1})\mathbf{M}(v_{2})\mathbf{\Phi }(0,\pi
/4,-3\pi /4) \\
\left\vert v_{1}\right\rangle &=&\frac{1}{2}\sqrt{1+\frac{1}{\sqrt{3}}}\left[
1-\sqrt{3},1,1\right] ^{T},  \label{V1-3} \\
\left\vert v_{2}\right\rangle &=&\sqrt{\frac{1+\sqrt{2}}{2\sqrt{2}}}\left[
0,1-\sqrt{2},\ -i\right] ^{T}.  \label{V2-3}
\end{eqnarray}%
The \emph{generalized}-QHR decomposition reads 
\end{subequations}
\begin{subequations}
\label{QFT3 generalized}
\begin{eqnarray}
\mathbf{U}_{3}^{F} &=&\mathbf{M}(v_{1};\pi )\mathbf{M}(v_{2};\pi /2), \\
\left\vert v_{1}\right\rangle &=&\frac{1}{2}\sqrt{1+\frac{1}{\sqrt{3}}}\left[
1-\sqrt{3},1,1\right] ^{T},  \label{V1-3g} \\
\left\vert v_{2}\right\rangle &=&\frac{1}{\sqrt{2}}\left[ 0,1,-1\right] ^{T}.
\label{V2-3g}
\end{eqnarray}%
Here the first QHR $\mathbf{M}(v_{1};\pi )=\mathbf{M}(v_{1})$ is the same
for the standard- and generalized-QHR implementations.

\subsection{Quartit}

For a quartit ($N=4$) the QFT matrix reads 
\end{subequations}
\begin{equation}
\mathbf{U}_{4}^{F}=\frac{1}{2}\left[ 
\begin{array}{cccc}
1 & 1 & 1 & 1 \\ 
1 & i & -1 & -i \\ 
1 & -1 & 1 & -1 \\ 
1 & -i & -1 & i%
\end{array}%
\right] .  \label{QFT4}
\end{equation}%
The \emph{standard}-QHR decomposition reads 
\begin{subequations}
\label{QF4 standard}
\begin{eqnarray}
\mathbf{U}_{4}^{F} &=&\mathbf{M}(v_{1})\mathbf{M}(v_{2})\mathbf{\Phi }(0,\pi
/4,0,-3\pi /4), \\
\left\vert v_{1}\right\rangle &=&\frac{1}{2}\left[ -1,1,1,1\right] ^{T}, \\
\left\vert v_{2}\right\rangle &=&\sqrt{\frac{1+\sqrt{2}}{2\sqrt{2}}}\left[
0,1-\sqrt{2},0,\ -i\right] ^{T}.
\end{eqnarray}%
The \emph{generalized}-QHR decomposition reads 
\end{subequations}
\begin{subequations}
\label{QF4 generalized}
\begin{eqnarray}
\mathbf{U}_{4}^{F} &=&\mathbf{M}(v_{1};\pi )\mathbf{M}(v_{2};\pi /2), \\
\left\vert v_{1}\right\rangle &=&\frac{1}{2}\left[ -1,1,1,1\right] ^{T},
\label{V1-4g} \\
\left\vert v_{2}\right\rangle &=&\frac{1}{\sqrt{2}}\left[ 0,1,0,-1\right]
^{T}.  \label{V2-4g}
\end{eqnarray}%
Again, the first QHR $\mathbf{M}(v_{1};\pi )=\mathbf{M}(v_{1})$ is the same
for the standard- and generalized-QHR implementations. Interestingly, the
QFT for $N=4$ is decomposed with only \emph{two} QHRs, rather than three, 
\emph{without} phase gates.

\begin{figure}[tbp]
\includegraphics[width=70mm]{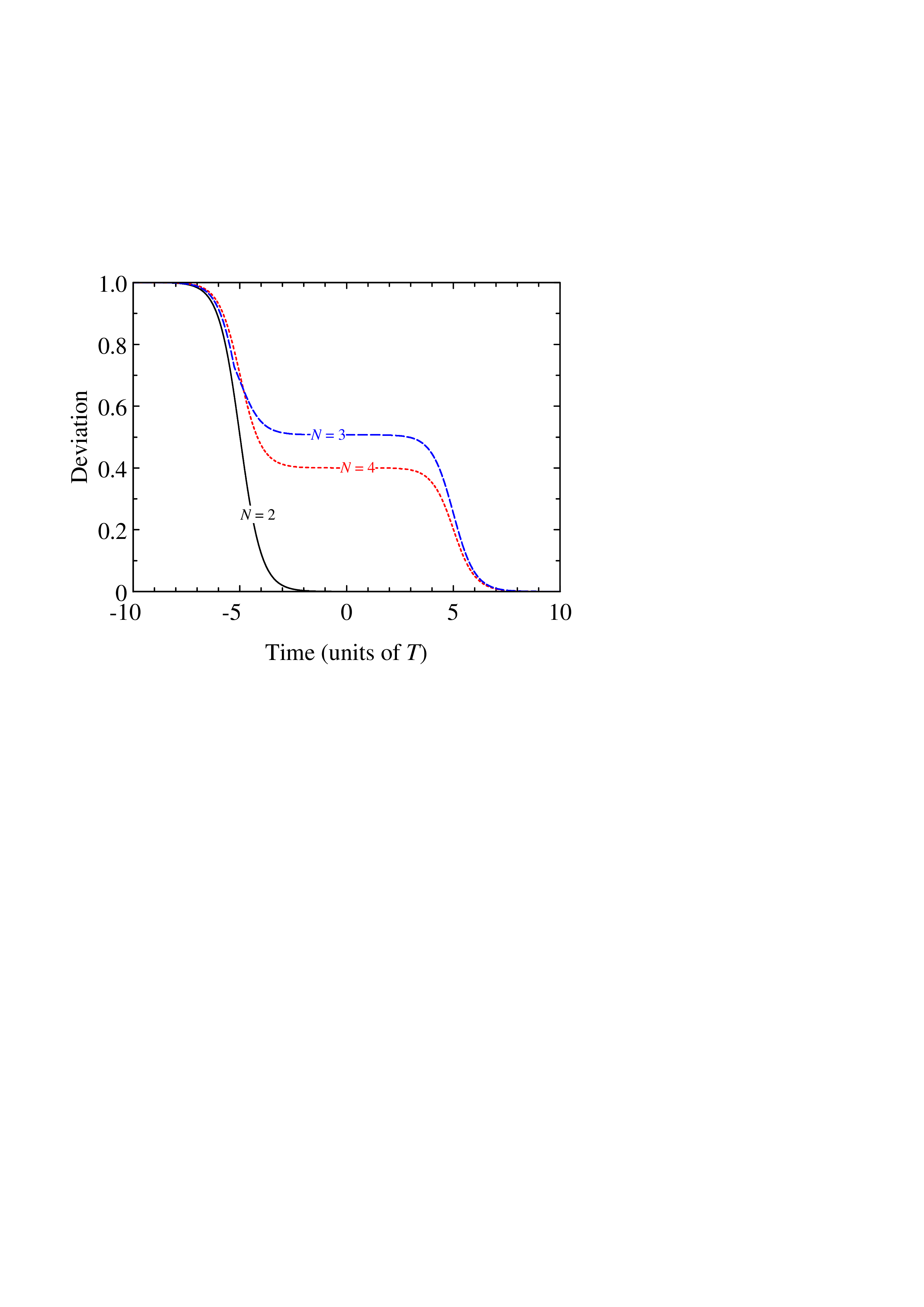}
\caption{Deviation $\sum_{j,k=1}^{N}\left\vert \left( U_{N}\right)
_{jk}-\left( U_{N}^{F}\right) _{jk}\right\vert $ of the propagator $\mathbf{U%
}_{N}(t)$ from the QFT matrix $\mathbf{U}_{N}^{F}$ versus time for $N=2,3,4$%
, for generalized-QHR implementations. The pulses for $N=2$ are centered at
time $\protect\tau =-5T$, whereas for $N=3$ and $4$ at times $\protect\tau %
=-5T$ and $\protect\tau =5T$. We have assumed sech pulse shapes (\protect\ref%
{pshape}) and rms pulse area $A=2\protect\pi $ ($\protect\chi =2$). The
individual couplings $\protect\chi _{n}$ are given by the components of the
generalized-QHR vectors (\protect\ref{V-2}) for $N=2$, (\protect\ref{V1-3g})
and (\protect\ref{V2-3g}) for $N=3$, (\protect\ref{V1-4g}) and (\protect\ref%
{V2-4g}) for $N=4$, each multiplied by $\protect\chi $. All phases $\protect%
\beta _{n}$ are zero. The detunings are $\Delta =0$ for the first steps and $%
\Delta =1/T$ for the second.}
\end{figure}

Figure \ref{Fig-QFT} shows the time evolution of the propagator $\mathbf{U}%
_{N}(t)$ towards the respective QFT matrix $\mathbf{U}_{N}^{F}$, for $%
N=2,3,4 $, for realizations with generalized QHRs. As time progresses, the
deviation of $\mathbf{U}_{N}(t)$ from $\mathbf{U}_{N}^{F}$ vanishes steadily
in all cases. As predicted, QFT is realized with just a single QHR for $N=2$
and with just two QHRs for $N=3$ and 4.

\section{Discussion and conclusions}

We have proposed a simple physical implementation of the quantum Householder
reflection in a coherently driven $N$-pod system. We have shown that the
most general U($N$) transformation of a qunit can be constructed by at most $%
N-1$ standard QHRs and an $N$-dimensional phase gate, or by $N-1$
generalized QHRs (each having an extra phase parameter compared to the
standard QHR) and a one-dimensional phase gate, i.e. by only $N$ physical
operations. This significant improvement over the existing setups [involving 
$O(N^{2})$ operations] can be crucial in making quantum state engineering
and operations with qunits experimentally feasible.

The Householder gate is superior already for a qubit because the general
U(2) gate needs just two gates, a QHR and a phase gate, compared to three
gates in existing implementations. For a qutrit, the QHR realization of U(3)
requires only three gates, compared to at least seven hitherto. The QHR
implementation of the U($N$) gate is particularly important for qutrits
because of the straightforward physical implementation in a $%
J=1\leftrightarrow J=0$ transition (Fig. \ref{Fig-Npod}); the results, of
course, apply to any $N$, and can be accomplished, for instance, by using
more $J=1$ levels.

We have given examples for QHR implementations of quantum Fourier
transforms. The QHR realization of QFT for a qubit requires a single
interaction step, compared to two steps hitherto. The QHR realization is
particularly efficient for a quartit ($N=4$), where the QFT is synthesized
with \emph{only two} QHR gates [as for a qutrit ($N=3$)], much fewer than $%
O(4^{2})$ in the existing SU(2) proposals. The components of the Householder
vectors are the amplitudes of the respective couplings. It is important that
all QHR phases are relative phases of the external control fields, e.g.
relative laser phases, which are much easier to control than dynamic and
geometric phases.

The generalized QHR requires off-resonant pulsed interactions, appropriately
detuned from resonance. The standard QHR can be realized both on and off
resonance. The off-resonance implementation has the advantage that only
negligible transient population is placed into the (possibly decaying)
ancillary excited state; however, it requires a specific value of the
detuning.

In the existing SU(2) proposals, each interaction step involves a single
SU(2) (or Givens) rotation. The difference between the Givens rotation and
the Householder reflection is that, when applied to an arbitrary matrix, the
Givens rotation nullifies a single matrix element; the Householder
reflection nullifies an entire row (or column). When the matrix is unitary,
a single Householder reflection nullifies one column and one row
simultaneously. Hence the Householder reflection is N times faster than the
Givens SU(2) transformation.

In atoms and ions qubits are encoded usually in degenerate ground sublevels,
and the coupling between them is accomplished by off-resonant interactions,
via an intermediate state, which is eliminated adiabatically to produce an
effective Raman coupling. In doing so, the phase relation between the two
Raman fields is lost. In our proposal we use resonant, or nearly-resonant,
fields; no adiabatic elimination is performed and the phase relation is
preserved in the resulting QHR propagator. Therefore, already for $N=2$, the
QHR contains an additional phase parameter compared to previous
realizations, which reduces the number of steps for U(2) operations from 3
to 2. Hence, even for a qubit there is a clear improvement. It is also
significant that resonant interactions, which we use, require less
interaction energy than off-resonant interactions; this may be crucial in
the case of weak couplings.

We conclude by emphasizing that the wide-spread use of the Householder
reflection in classical data analysis promises that the proposed quantum
implementation has the potential to become a powerful tool for quantum state
engineering and quantum information processing.

\acknowledgments

This work is supported by the European Union's ToK project CAMEL, MCTS
project BERFLE, RTN project EMALI, and the Alexander von Humboldt Foundation.

\end{subequations}

\end{document}